\documentclass[12pt, draftclsnofoot, onecolumn]{IEEEtran}
\IEEEoverridecommandlockouts
\usepackage{booktabs}
\usepackage{multirow}
\usepackage{tabularx}
\usepackage{cite}
\usepackage{amsmath,amssymb,amsfonts}
\usepackage{algorithmic}
\usepackage{graphicx}
\usepackage{textcomp}
\usepackage{balance}
\usepackage{epsfig,latexsym}
\usepackage{cite}
\usepackage{subfigure}
\usepackage[ruled,vlined]{algorithm2e}
\usepackage{stfloats}
\usepackage{epstopdf}
\usepackage{bm}
\usepackage[justification=centering]{caption}
\usepackage{xcolor}
\usepackage{epsfig,latexsym}
\usepackage{caption}
\usepackage{verbatim}
\def\BibTeX{{\rm B\kern-.05em{\sc i\kern-.025em b}\kern-.08em
		T\kern-.1667em\lower.7ex\hbox{E}\kern-.125emX}}
\begin{document}

\title{AI-enabled STAR-RIS aided MISO ISAC Secure Communications\\
\thanks{}
}

\author{\IEEEauthorblockN{Zhengyu Zhu,~\emph{Senior Member, IEEE}, Mengfei Gong, Gangcan Sun, Peijia Liu, De Mi}
		\vspace{-0.19in}

\thanks{Z. Zhu is with the School of Electrical and Information Engineering, Zhengzhou University, Zhengzhou, 450001, China, and is also with National Mobile Communications Research Laboratory, Southeast University, Nanjing
	210019, China. (e-mail: iezyzhu@zzu.edu.cn).
	
	M. Gong, G. Sun and Peijia Liu are with the School of Electrical and Information Engineering, Zhengzhou University,
   Zhengzhou 450001, China. (e-mail: gmf2024@gs.zzu.edu.cn, iegcsun@zzu.edu.cn, peijialiu\_academic@yeah.net).
	
	D. Mi is with the School of Computing and Digital Technology, Birmingham City University, UK (e-mail: de.mi@bcu.ac.uk).
	}		}

\maketitle
\begin{abstract}
A simultaneous transmitting and reflecting reconfigurable intelligent surface (STAR-RIS) aided integrated sensing and communication (ISAC) dual-secure communication system is studied in this paper. The sensed target and legitimate users (LUs) are situated on the opposite sides of the STAR-RIS, and the energy splitting and time switching protocols are applied in the STAR-RIS, respectively. The long-term average security rate for LUs is maximised by the joint design of the base station (BS) transmit beamforming and receive filter, along with the STAR-RIS transmitting and reflecting coefficients, under guaranting the echo signal-to-noise ratio thresholds and rate constraints for the LUs. Since the channel information changes over time, conventional convex optimization techniques cannot provide the optimal performance for the system, and result in excessively high computational complexity in the exploration of the long-term gains for the system. Taking continuity control decisions into account, the deep deterministic policy gradient and soft actor-critic algorithms based on off-policy are applied to address the complex non-convex problem. Simulation results comprehensively evaluate the performance of the proposed two reinforcement learning algorithms and demonstrate that STAR-RIS is remarkably better than the two benchmarks in the ISAC system.

\end{abstract}


\section{Introduction}
The radio spectrum is becoming increasingly overcrowded with the rapid growth of connected devices and mobile applications. The majority of commercial communication networks have traditionally worked in the sub-6 GHz band, which coexist peacefully with the existing radar sensing systems that are used for such applications as air traffic regulation and weather forecasting\cite{005}. However, the increasing demand for wireless connections is driving the expansion of wireless communications into higher frequency bands such as millimeter wave and terahertz, which is causing an increasing overlap with the traditional radar sensing bands\cite{006}. The continued independent design of communications and radar sensing systems would cause severe coexistence interference\cite{007}. Thus, it is necessary to jointly design the communication and sensing systems for them share the spectrum in ordor to reduce interference and improve spectrum efficiency. Furthermore the difference between wireless communication and radar sensing is gradually becoming narrow with the development of massive multiple-input multiple-output (MIMO), making it feasible to jointly design. Morever, the 6G network is expected to achieve excellent wireless communication and simultaneous highly precise radio sensing, which are vital for the emerging applications, including smart cities and internet of vechicles. Hence, the integrated sensing and communication (ISAC) has recently become a mainstream trend and a hot research issue in the context of 6G technology\cite{008,009,013}.

ISAC is a novel information processing technology that can effectively improve the spectral, information processing and hardware efficiency for the systems via sharing hardware resources and information to achieve the coordination between sensing and communication\cite{011,012}. With the methods of time-frequency space resource multiplexing, hardware equipment sharing, and the joint design of air-interface and protocols, ISAC is able to provide a unified management for communication and sensing, and also provide the support for both communication and sensing independently in the same system. However, in practical applications, ISAC systems face a range of challenges, such as the high environment dependence, limited coverage, high cost and high power consumption constraints. This is because high frequency transmission result in high pathloss, non-line-of-sight (NLoS) path signals are usually weak, while the line-of-sight (Los) path may be blocked by the environmental obstacle, resulting in blind areas\cite{014,015,016}. And these environmented factors may impose big impacts on the performance of ISAC systems.

On the other side, the development of reconfigurable intelligent surfaces (RIS), which are artificial surfaces equipped with an array of low-cost passive components, has provided an innovative strategy to overcome the challenges faced by ISAC systems\cite{004}. Specifically , the intelligent controller of RIS regulates the phase response for every component to manage the propagation of wireless signals incident on the RIS, which can help to break through the uncertainty of the conventional wireless transmission environments, making the propagation environments feasible for implementfation of ISAC \cite{001}. Nevertheless, the conventional RIS is designed to only reflect the incident signal, providing a 180-degree semi-plane reflection, and the served users must be situated on the same side of both RIS and base station (BS), resulting in the relatively small coverage. To offset this restriction, the simultaneous transmitting and reconfigurable intelligent surface (STAR-RIS) has been proposed to obtain 360-degree coverage, which can serve the users located anywhere around the STAR-RIS\cite{002,003,010}. With this regard, when STAR-RIS is integrated with ISAC, it will not only expand the coverage of high frequency band systems, but also enhance the communication and sensing performance, while reducing system's budget and power consumption. Furthermore, their joint will lead to more possibilities for the design in the wireless communication and sensing fields.

\subsection{Related Works} 
So far, the RIS-aidded ISAC systems have been explored in some reference \cite{101,102,103,104}. Sepcifically in \cite{101}, the RIS-assisted dual-functional radar and communication (DRFC) system is studied, where a multi-antenna BS utilizes the same hardware platform to perform simultaneous multi-user communication and radar sensing. The state-of-the-art of the RIS-aidded DRFC scheme is validated for maximizing the signal-to-noise ratio (SNR) at radar output, via jointly designing the beamformers for both the BS dual-functional transmitter and RIS. A RIS-aidded ISAC system operated in the millimeter-wave band is presented in \cite{102}. The joint design of the covariance matrix for radar signaling, beamwave of communication system, and the RIS phase shift is studied, with the objective to achieves the communication rate improvement, while fulfilling the desired radar waveforms. It is shown that RIS is capable of effectively enhancing the performance of the ISAC system. The authors in \cite{103} studied the double RIS-aided ISAC system, which uses two RISs to jointly optimize the RIS and radar beamforming so as to reinforce the communication signals and inhibit each other's interferences under the radar detection constraints. Also, a favorable discussion of the validity of using multiple RISs to assist ISAC is offered. In \cite{104}, the joint design is studied based on the practical consideration of the constant-mode waveform and the discrete RIS phaseshifts. The studies show that it is possible to minimize the inter-user interference under the Clamey-Rao constraint on the target direction-of-arrival (DOA) estimation. Overally, the studies as above mentioned have demonstrated the effectiveness of the RIS-aided ISAC in practice.  


STAR-RIS has also been widely studied due to its high degrees of freedom for implementation. In \cite{105}, the authors proposed three operational protocols, namely, time switching (TS), energy splitting (ES) and mode switching (MS), in the context of the fundamental STAR-RIS signaling model. The performance of the above-mentioned protocols with unicast and multicast transmission respectively has been studied, which provides the helpful guideline for designing the STAR-RIS assisted radio communication systems. The STAR-RIS assisted ISAC systems has also been addressed in \cite{106}, when both the individual and coupled phaseshift STAR-RIS models are considered. Through minimizing the DOA estimation of being sensed targets under the constraints of the Clamello boundaries and communication requirements, it is shown that it is feasible to adopt STAR-RIS to simultaneously enhance the quality of sensing and the performence of communications.

\subsection{Motivations and Contributions}
While the research on the STAR-RIS aided ISAC system has just been started, the security of the STAR-IRS-aided ISAC systems have not received much attention. Furthermore, in the open literature, no works can be found on the synergy existing between STAR-RIS and ISAC, especially, when relatively long-term time is considered. This is bacause channel changes over time due to the mobility of both users and targets. Hence, it is necessary for the proposed algorithms to be robust to the channel variations when over a relatively long period. In this case, the conventional convex optimization techniques may not  provide the optimal performance, while they usually result in excessively high computational complexity.

In against the background in practically, we motivate to develop coupled STAR-RIS model emploied by both TS and ES operating protocols so as to guarantee the secure transmission in ISAC systems. On the basis of the proposed model, the secure communication for legitimate users is investigated and optimised. Overall, the following contributions are made in this paper.
\begin{enumerate}
	\item We study the joint STAR-RIS and ISAC secure communication systems, in which the STAR-RIS utilises a coupled phase-shift model and employs two operating protocols, namely ES and TS, respectively. Our objective is to prevent information leakage from multiple legitimate users (LUs) to the potential eavesdroppers,  i.e. sensing target (ST) and eavesdroppers (Eve). The secrecy performence of LUs over long-time period is maximised via designing the transmit beamforming and receive filters for BS, along with reflecting and transmitting coefficients.
	\item However, the optimization of STAR-RIS problem for long time periods and large-scale STAR-RIS is ectremely high complexity, which is hard to be solved due to the STAR-RIS coupled phase-shift model. To tackle this problem, we propose two joint design algorithms based on, respectively, the deep deterministic policy gradient (DDPG) and soft actor-critic (SAC) to optimize the discrete and continuous hybrid actions.
	\item Our studies and results show that the SAC network has a better exploration capability and higher learning efficiency, but  is more complex than the DDPG network. The STAR-RIS performs better than both the dual splicing RIS and conventional RIS. Furthermore, the TS protocol for STAR-RIS is better than the ES protocol in the STAR-RIS assisted ISAC systems.
\end{enumerate}

\textbf{Notations}: Uppercase boldface and lowercase boldface represent matrix and vector, respectively. ${\text{Tr}}\left( {\mathbf{E}} \right)$ denotes the trace of a matrix ${\mathbf{E}}$. ${{\bf{E}}^H}$ stands for the conjugate transpose operations of ${\mathbf{E}}$. ${\text{diag}}\left( {\mathbf{e}} \right)$ returns a diagonal matrix. $|\cdot|$ and $|| \cdot ||$ indicate the absolute value of a complex number and Euclidean norm of a vector. $\left\lfloor {} \right\rfloor$ define as the bottom value function. $[\cdot]^{+}=\max \{0, \cdot\}$. 

\section{SYSTEM MODEL AND PROBLEM FORMULATION}
\subsection{System Model}
We consider a STAR-RIS-assisted ISAC system containing $M$ LUs, an EVE and a ST with single antenna as shown in Fig. 1. The system involves a multi-antenna BS equipped with a uniform linear array comprising $L_1$ transmit antennas and $L_2$ receive antennas, and a STAR-RIS array with $N$ passive transmit-reflect units arranged in a uniform planar array \cite{200}. Without any lose of generality, we set $L_1 = L_2 = L$ for simplicity. The employment of STAR-RIS leads the entire space to be divided into a sensing space at the reflecting side and a communication space at the transmitting side\cite{201}. The BS executes target sensing while transmitting information to LUs. Nevertheless, the existence of perception target and Eve could eavesdrop on the LUs' information. It is crucial to ensure that the LU's information remains confidential and is not compromised by either the ST or Eve.
\begin{figure}[ht]
	\centerline{\includegraphics[width=0.7\textwidth]{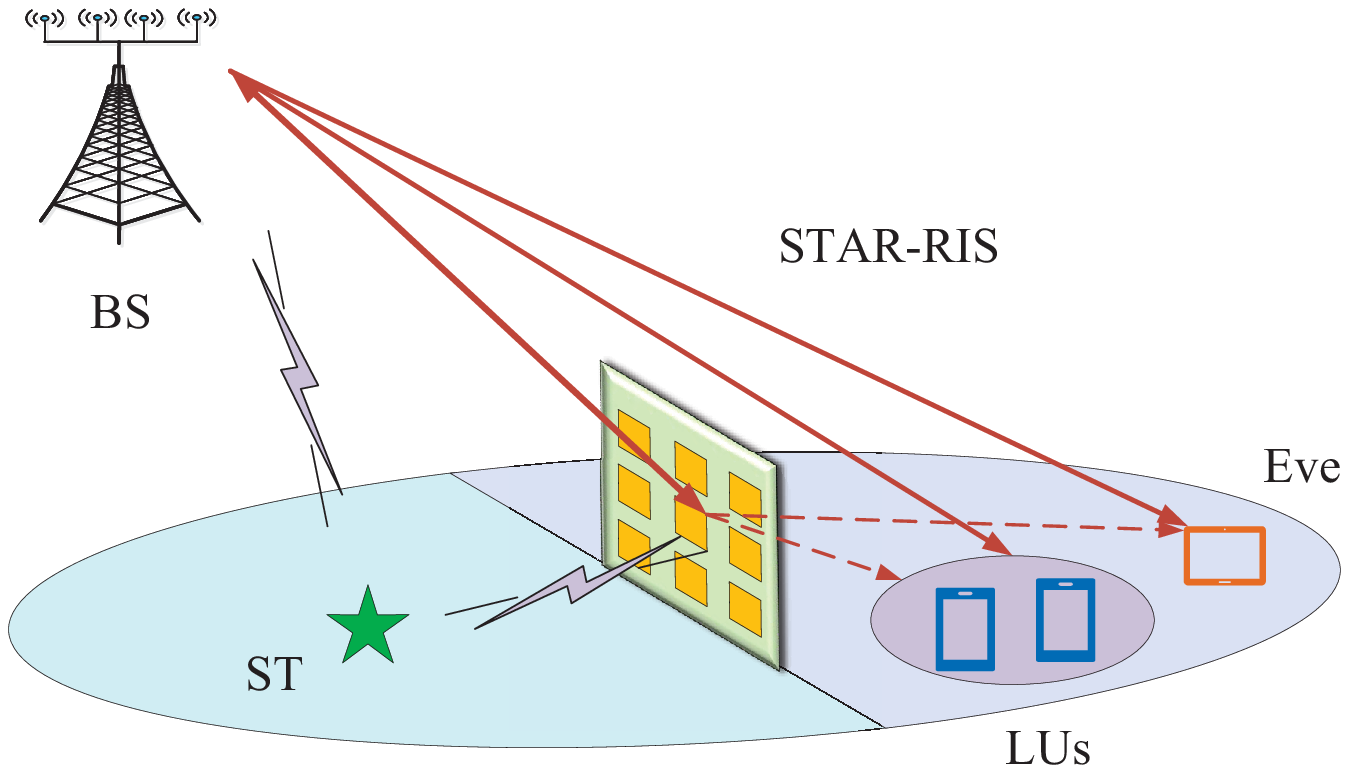}}
	\caption{The ISAC secure communication system enhanced by STAR-RIS}
	\label{fig}
\end{figure}

\subsection{STAR-IRS Model}
In this subsection, we firstly discuss the ES model of STAR-RIS for handling the incident signals by classifying them into reflected signals for the perception of $A$ and transmitting signals for the communication of $B$. Hence, the STAR-RIS reflected and transmitted phase-shift matrix are respectively denoted as  
\begin{equation}
	{\bf{\Phi}} _i = diag\left( {\alpha _1^i{e^{j\varphi _1^i}}, \ldots ,\alpha _N^i{e^{j\varphi _N^i}}} \right),\forall i \in \left[ {A,B} \right],
\end{equation}
where $\varphi _n^i \in [ - \pi$,$\pi ]$ and $\alpha _n^i \in [0,1]$ are respectively the phase shift and amplitude of the \emph{n}-th unit for STAR-RIS\cite{1001}.
In practice, it is challenging to independently adjust the reflective and transmissive factors due to the fact that these values are determined by resistance and reactance. Thus we make use of the coupling relation for amplitude as\cite{202} 
\begin{equation} 
	{\left( {\alpha _n^A} \right)^2} + {\left( {\alpha _n^B} \right)^2} = 1,\forall n \in \left[ {1,N} \right].
\end{equation}
In addition, a coupling phase shift is given by
\begin{equation} 
	\cos \left( {\varphi _n^A - \varphi _n^B} \right) = 0,\forall n \in \left[ {1,N} \right].
\end{equation}

Then, the TS model of STAR-RIS, which is different from ES, switches between single reflection and single transmission mode periodically for every element during different orthogonal timeslots. Accordingly, $ \pi_1 $ and $ \pi_2 $ are defined as the reflection and transmission time as a ratio of the overall timeslot, respectively, where $0 \leq \pi_1 \leq 1$, $0 \leq \pi_2 \leq 1$ and $\pi_1 + \pi_2 = 1$. This means that STAR-RIS reflects signals with $ \pi_1 $, and transmit signals with $ \pi_2 $ of the total time. Thus, the reflection and transmission phase shifts of STAR-RIS in TS mode can be positioned as ${\Phi _A^{TS}} = diag\left( {{e^{j\varphi _1^A}}, \ldots ,{e^{j\varphi _N^A}}} \right)$ and
${\Phi _B^{TS}} = diag\left( {{e^{j\varphi _1^B}}, \ldots ,{e^{j\varphi _N^B}}} \right)$, where $\varphi _n^i \in [ 0$, $2\pi )$. Since the transmission and reflection coefficients in TS mode are uncoupled in the time domain, it is relatively easily to optimize. However, switching between two phases periodically makes strict demand on time synchronization, which may be challenging in practical implementation. 

\subsection{Channel Model}
Assume that in a period $\emph{T}$, all channel coefficients experience the quasi-static flattened fading. Let $\lambda _{b,r,t}$ denote the channel loss from BS to STAR-RIS in the \emph{t}-th time slot. Futhermore, assume the Racian channel model. Then, the channel matrix can be represtented as
\begin{equation} 
	{{\bf{H}}_t} = \sqrt {{\lambda _{b,r,t}}} \left( {\sqrt {\frac{F}{{F + 1}}} {{\bf{u}}_{b,r,t}} + \sqrt {\frac{1}{{F + 1}}} {{\bf{v}}_{b,r,t}}} \right),
\end{equation}
where ${\bf{v}}_{b,r,t}$ and ${{\bf{u}}_{b,r,t}} = {f _r}\left( {{\beta _r},{\zeta _r}} \right)f _b^T\left( {{\beta _b}} \right)$ are the NLoS and LoS components, and $F$ stands for the Rician factor. Furthermore, $f _r \in {\mathbb{C}^{N \times 1}}$ is defined as the AOD at BS, while $f _b \in {\mathbb{C}^{M \times 1}}$ and $\psi _r$ are the direction angle and elevation angle arriving at STAR-RIS. Based on \cite{203}, assume that ${{N_x}}$ stands for the total number of STAR-RIS elements in each row, ${{d_r}}$ and ${{d_0}}$ mean the distances between the neighboring reflective elements and neighboring antennas, respectively. Then we have ${f _r}\left[ n \right] = {e^{j2\pi {d_r}(n - 1)\frac{{\left( {\left\lfloor {n/{N_s}} \right\rfloor {\eta _1}{e^{\left( {n - \left\lfloor {n/{N_s}} \right\rfloor {N_s}} \right){\eta _2}}}} \right)}}{\lambda }}}$ and ${f _b}\left( l \right) = {e^{\frac{{j2\pi \left( {l - 1} \right){d_0}\sin \left( {{\varphi _b}} \right)}}{\lambda }}}$, where ${\eta _1} = \sin \left( {{\beta _r}} \right)\sin \left( {{\zeta _r}} \right)$ and ${\eta _2} = \sin \left( {{\beta _r}} \right)\cos \left( {{\zeta _r}} \right)$.

It is assumed that BS has perfect channel state information (CSI) and that the path follows an urban transmission loss model\cite{204}. Then the path loss of the LoS component is ${\lambda _{LoS}}\left( {d,{f_1}} \right) = 20{\log _{10}}{f_1} + 22.0{\log _{10}}d + 28.0$, and that of the NLoS component is ${\lambda _{NLoS}} = \max \left[ {{\lambda _{NLoS}}\left( {d,{f_1}} \right),{\lambda _{LoS}}\left( {d,{f_1}} \right)} \right]$, where ${\lambda _{NLoS}}\left( {d,{f_1}} \right) = 26{\log _{10}}{f_1} + 36.7{\log _{10}}d + 22.7 - 0.3\left( {{z_r} - 1.5} \right)$. Accordingly, the channels from the BS and STAR-RIS to LU, Eve and LU are denoted as, ${{\bf{h}}_{b,m,t}} = \sqrt {{\lambda _{b,m,t}}} {{\bf{v}}_{b,m,t}}$, ${{\bf{h}}_{r,m,t}} = \sqrt {{\lambda _{r,m,t}}} {{\bf{v}}_{r,m,t}}$, ${{\bf{h}}_{b,e,t}} = \sqrt {{\lambda _{b,e,t}}} {{\bf{v}}_{b,e,t}}$, ${{\bf{h}}_{r,e,t}} = \sqrt {{\lambda _{r,e,t}}} {{\bf{v}}_{r,e,t}}$, ${{\bf{g}}_{b,s,t}} = \sqrt {{\lambda _{b,s,t}}} {{\bf{v}}_{b,s,t}}$, and ${{\bf{g}}_{r,s,t}} = \sqrt {{\lambda _{r,s,t}}} {{\bf{v}}_{r,s,t}}$, respectively.

\subsection{Signal Model}
\textit{Energy switching model}: Assume that the beamforming vectors and signals for communication and radar signals be given as ${\bf{k}}_s$ and ${c_s}$, and ${\bf{k}}_w$ and ${c_w}$, respectively. Then, the signals transmitted by the BS in timeslot $t$ can be written as 
\begin{equation}
{\bf{c}}\left( t \right) = {{\bf{k}}_s}\left( t \right){c_s}\left( t \right) + {{\bf{k}}_w}\left( t \right){c_w}\left( t \right). 
\end{equation}
For convenience of presentation, we represent the symbol vector as 
$\mathbf{c}[t] \triangleq\left[\mathbf{c}_{\mathrm{s}}^{T}[t] \quad \mathbf{c}_{\mathrm{w}}^{T}[t]\right]^{T}$ and the beamforming matrix as $\mathbf{k}\left( t \right) \triangleq\left[\begin{array}{ll}\mathbf{k}_{\mathrm{s}}\left( t \right) & \mathbf{k}_{\mathrm{w}}\left( t \right)\end{array}\right]$.
The LU received signal in the \emph{t}-th timeslot can be expressed as 
\begin{equation} 
	{y_{m,t}} = {\bf{h}}_{m,t}^H{{\bf{k}}_s}\left( t \right){c_s}\left( t \right) + \sum\limits_{i = 1}^L {{\bf{h}}_{m,t}^H{{\bf{k}}_{w,i}}\left( t \right){c_{w,i}}} \left( t \right) + {n_m}\left( t \right), m \in M,
\end{equation}
where ${\bf{h}}_{m,t}^H = \left( {{\bf{h}}_{r,m,t}^H{\Phi _{B,t}}{{\bf{H}}_t} + {\bf{h}}_{b,m,t}^H} \right)$ and ${n_m}\left( t \right) \sim CN\left( {0,\sigma _m^2} \right)$ stands for the additive white gaussian noise (AWGN). Accordingly, the signal-to-interference-plus-noise ratio (SINR) of the \emph{m}-th LU is given by
\begin{equation}
	{\bf{SINR}}_{m,t} = \frac{{{{\left| {{\bf{h}}_{m,t}^H{{\bf{k}}_s}\left( t \right)} \right|}^2}}}{{\sum\limits_{i = 1,i\ne m}^M {{{\left| {{\bf{h}}_{m,t}^H{{\bf{k}}_{s,i}}\left( t \right)} \right|}^2}} + \sum\limits_{j = 1}^L {{{\left| {{\bf{h}}_{m,t}^H{{\bf{k}}_{w,j}}\left( t \right)} \right|}^2}} + \sigma _m^2}}, m \in M.
\end{equation}

Similarly, the SINR for Eve and ST to steal the information of the \emph{m}-th LU can be formulated as
\begin{equation}
	{\bf{SINR}}_{m,t}^e = \frac{{{{\left| {{\bf{h}}_{e,t}^H{{\bf{k}}_s}\left( t \right)} \right|}^2}}}{{\sum\limits_{i = 1,i\ne m}^M {{{\left| {{\bf{h}}_{e,t}^H{{\bf{k}}_{s,i}}\left( t \right)} \right|}^2}} +\sum\limits_{i = 1}^L {{{\left| {{\bf{h}}_{e,t}^H{{\bf{k}}_{w,i}}\left( t \right)} \right|}^2}}  + \sigma _e^2}},\quad m \in M,
\end{equation}
\begin{equation}
	{\bf{SINR}}_{m,t}^s = \frac{{{{\left| {{\bf{g}}_{s,t}^H{{\bf{k}}_s}\left( t \right)} \right|}^2}}}{{\sum\limits_{i = 1,i\ne m}^M {{{\left| {{\bf{g}}_{s,t}^H{{\bf{k}}_{s,i}}\left( t \right)} \right|}^2}} +\sum\limits_{i = 1}^L {{{\left| {{\bf{g}}_{s,t}^H{{\bf{k}}_{w,i}}\left( t \right)} \right|}^2}}  + \sigma _s^2}},\quad m \in M,
\end{equation}
where ${\bf{h}}_{e,t}^H = {\bf{h}}_{r,e,t}^H{\bf{\Phi} _{B,t}}{{\bf{H}}_t} + {\bf{h}}_{b,e,t}^H$ and ${\bf{g}}_{s,t}^H = {\bf{g}}_{r,s,t}^H{\bf{\Phi} _{A,t}}{{\bf{H}}_t} + {\bf{g}}_{b,s,t}^H$.

To ensure the information security for all LUs, we have to consider the existence of both Eve and ST to formulate the sum secrecy rate for the LU $m$ in the timeslot $t$ as\cite{205,1000,1002,1003}
\begin{equation}
	R_{m,t}^{\sec } = {\left[ {{R_{m,t}} - R_{m,t}^e} \right]^ + } + {\left[ {{R_{m,t}} - R_{m,t}^s} \right]^ + },
\end{equation}
where $R_{m,t}^s = \log \left( {1 + {\bf{SINR}}_{m,t}^s} \right)$, $R_{m,t}^e = \log \left( {1 + {\bf{SINR}}_{m,t}^e} \right)$ and ${R_{m,t}} = \log \left( {1 + {\bf{SINR}}_{m,t}} \right)$.

At the same time, the compound magnitude is given as ${\tau _t}$ with ${\rm E}\left\{ {{{\left| {{\tau _t}} \right|}^2}} \right\} = \sigma _t^2$. Then the received echo signal of BS is
\begin{equation}
	{y_s}\left[ t \right] = {\tau _t}{{\bf{g}}_{s,t}}{\bf{g}}_{s,t}^H{\bf{c}}\left[ t \right] + {n_s}\left[ t \right].
\end{equation}  

After the matched filtering, the signals received over $P$ time slots can be written as 
\begin{equation}
	{{\bf{Y}}_s} = {\tau _t}{{\bf{H}}_s}{\bf{kC}}{{\bf{C}}^H} + {{\bf{N}}_s}{{\bf{C}}^H},
\end{equation}
where ${\bf{H}}_s = {{\bf{g}}_{s,t}}{\bf{g}}_{s,t}^H$, ${\bf{C}} \buildrel \Delta \over = \left[ {C\left[ 1 \right],...,C\left[ P \right]} \right]$, and ${{\bf{N}}_s} \buildrel \Delta \over = \left[ {{n_s}\left[ 1 \right],...,{n_s}\left[ P \right]} \right]$. In addition, the receive filter ${\bf{u}} \in {\mathbb{C}^{L \times (1+L)}}$ is utilized to amplify the echo signal
\begin{equation}
	{{\bf{u}}^H}{{\bf{\tilde y}}_{\bf{s}}} = {\tau _t}{{\bf{u}}^H}\left( {{\bf{C}}{{\bf{C}}^H} \otimes {{\bf{H}}_s}} \right){\bf{\tilde k}} + {{\bf{u}}^H}{{\bf{\tilde n}}_{\bf{s}}}.
\end{equation}
where ${ \bf{\tilde y}_{s}} \buildrel \Delta \over = vec\left\{ {{{\bf{Y}}_s}} \right\}$, ${ \bf{\tilde k}} \buildrel \Delta \over = vec\left\{ {\bf{k}} \right\}$ and ${{{\bf{\tilde n}}_{{s}}}} \buildrel \Delta \over = vec\left\{ {{{\bf{N}}_s}{{\bf{C}}^H}} \right\}$.

The SNR for the echo received by the BS is obtained by
\begin{equation}
	{\bf{SNR}} = \frac{{\tau _t^2E\left\{ {{{\left| {{{\bf{u}}^H}\left( {{\bf{C}}{{\bf{C}}^H} \otimes {{\bf{H}}_s}} \right){\bf{k}}} \right|}^2}} \right\}}}{{P\sigma _s^2{{\bf{u}}^H}{\bf{u}}}}.
\end{equation}
By the utilization of $E\left\{ {{\bf{C}}{{\bf{C}}^H}} \right\} = P{{\bf{I}}_{L + M}}$ and Jensen's inequality\cite{206}, the lower limit of the SNR can be acquired as
\begin{equation}
	{\bf{SN}}{{\bf{R}}} \ge {\bf{SN}}{{\bf{R}}_{0 }} = \frac{{P\tau _t^2{{\left| {{{\bf{u}}^H}\left( {{{\bf{I}}_{L+M}} \otimes {{\bf{H}}_s}} \right){\bf{k}}} \right|}^2}}}{{\sigma _s^2{{\bf{u}}^H}{\bf{u}}}}.
\end{equation}

Our aim is to maximize the sum secrecy rate for each LU in the entire period $\emph{T}$ via optimization of $\bf{k}$, $\Phi _A$, $\Phi _B$ and $\bf{u}$, while ensuring the achievable rate of each LU and echo SNR of BS.
Hence, the optimization problem can be formulated as
\begin{subequations}
	\begin{align}
		\mathop {\max }\limits_{{{\bf{k}},{{\bf{\Phi }}_A},{{\bf{\Phi }}_B},{\bf{u}}} } \quad& \sum\limits_{t = 1}^T \sum\limits_{m = 1}^M {R_{m,t}^{\sec }} \\
		s.t. \quad & E\left( {{\bf{k}}\left( t \right){{\bf{k}}^H}\left( t \right)} \right) \le {P_{0 }},\forall t, \\
		&{R_{m,t}} \ge {R_{0 }},\forall t, \forall m,\\
		&{\bf{SNR}}_{0} \ge {\kappa _t},\\
		&(1), (2), (3),
	\end{align}
\end{subequations}
where ${P_{0}}$ stands for the maximum transmit power of BS, ${R_{0 }}$ is the minimum required achievable rate for all LU, and ${\kappa _t}$ indicates the minimum echo SNR that can be detected by BS. For the problem (16), (16b) and (16c) are the BS transmit power and achievable rate of LU constraints, (16d) and (16e) represent the echo SNR detected by BS and STAR-RIS phaseshift constraints. Obviously, due to the nonconvex objective function (16a) and the nonconvex constraints (16b), (16c), (16d), (16e), as well as the entire period $T$, (16) is a non-convex problem, which is hard to solve. However, it is interesting to observe that the variable $\bf{u}$ is dependent only by the constraint (16d). With other parameters fixed, the updating for $\bf{u}$ is a simple Rayleigh quotient problem that can be given by
\begin{equation}
	{{\bf{u}}^*} = \frac{{\left( {{{\bf{I}}_{M + L}} \otimes {{\bf{H}}_s}} \right){\bf{\tilde k}}}}{{{{\bf{\tilde k}}^H}\left( {{{\bf{I}}_{M + L}} \otimes {\bf{H}}_s^H{{\bf{H}}_s}} \right){\bf{\tilde k}}}}.
\end{equation}

\textit{Time switching model}: 
The SINR of the \emph{m}-th LU in the \emph{t}-th timeslot in reflection period and transmission period can respectively formulated as

\begin{equation}
	{\bf{SIN}}{{\bf{R}}_{A,m,t}} = \frac{{{{\left| {{\bf{h}}_{b,m,t}^H{{\bf{k}}_s}\left( t \right)} \right|}^2}}}{{\sum\limits_{i = 1,i \ne m}^M {{{\left| {{\bf{h}}_{b,m,t}^H{{\bf{k}}_{s,i}}\left( t \right)} \right|}^2}}  + \sum\limits_{j = 1}^L {{{\left| {{\bf{h}}_{b,m,t}^H{{\bf{k}}_{w,j}}\left( t \right)} \right|}^2}}  + \sigma _m^2}},
\end{equation}

\begin{equation} 
	{\bf{SIN}}{{\bf{R}}_{B,m,t}} = \frac{{{{\left| {({\bf{h}}_{r,m,t}^H\Phi _{B,t}^{TS}{{\bf{H}}_t} + {\bf{h}}_{b,m,t}^H){{\bf{k}}_s}\left( t \right)} \right|}^2}}}{{\sum\limits_{i = 1,i \ne m}^M {{{\left| {({\bf{h}}_{r,m,t}^H\Phi _{B,t}^{TS}{{\bf{H}}_t} + {\bf{h}}_{b,m,t}^H){{\bf{k}}_{s,i}}\left( t \right)} \right|}^2}}  + \sum\limits_{j = 1}^L {{{\left| {({\bf{h}}_{r,m,t}^H\Phi _{B,t}^{TS}{{\bf{H}}_t} + {\bf{h}}_{b,m,t}^H){{\bf{k}}_{w,j}}\left( t \right)} \right|}^2}}  + \sigma _m^2}}.
\end{equation}

Accordingly, the achievable rate for the \emph{m}-th LU is given by
\begin{equation}
R_{m,t}^{TS} = {\pi _1}\log \left( {1 + {\bf{SINR}}_{A,m,t}} \right) + {\pi _2}\log \left( {1 + {\bf{SINR}}_{B,m,t}} \right).
\end{equation}

Similarly, the SINR for Eve and ST to steal the information of the \emph{m}-th LU in the \emph{t}-th timeslot is written as 
\begin{equation}
	R_{m,t}^{TS,e} = {\pi _1}\log \left( {1 + {\bf{SINR}}_{A,m,t}^e} \right) + {\pi _2}\log \left( {1 + {\bf{SINR}}_{B,m,t}^e} \right),
\end{equation}
\begin{equation}
	R_{m,t}^{TS,s} = {\pi _1}\log \left( {1 + {\bf{SINR}}_{A,m,t}^s} \right) + {\pi _2}\log \left( {1 + {\bf{SINR}}_{B,m,t}^s} \right).
\end{equation}

Hence, the sum secrecy rate for the \emph{m}-th LU in the \emph{t}-th timeslot is denoted as
\begin{equation}
	R_{m,t}^{TS,\sec } = {\left[ {{R_{m,t}^{TS}} - R_{m,t}^{TS,e}} \right]^ + } + {\left[ {{R_{m,t}^{TS}} - R_{m,t}^{TS,s}} \right]^ + }.
\end{equation}

Futhurmore, the upper threshold of echo SNR is expressed as
\begin{equation}
{\bf{SN}}{{\bf{R}}_{0 }^{TS}} = {\pi _1}\frac{{P\tau _t^2{{\left| {{{\bf{u_1}}^H}\left( {{{\bf{I}}_{K+L}} \otimes {{\bf{g}}_{b,s,t} {\bf{g}}_{b,s,t}^H)}} \right){\bf{k}}} \right|}^2}}}{{\sigma _s^2{{\bf{u_1}}^H}{\bf{u_1}}}}+{\pi _2}\frac{{P\tau _t^2{{\left| {{{\bf{u_2}}^H}\left( {{{\bf{I}}_{K+L}} \otimes {{\bf{g}}_{s,t}^{TS}({\bf{g}}_{s,t}^{TS})^H}} \right){\bf{k}}} \right|}^2}}}{{\sigma _s^2{{\bf{u_2}}^H}{\bf{u_2}}}}.
\end{equation}
where $({\bf{g}}_{s,t}^{TS})^H = {\bf{g}}_{r,s,t}^H{\Phi _{A,t}^{TS}}{{\bf{H}}_t} + {\bf{g}}_{b,s,t}^H$.

Therefore, in the TS model, the sum secrecy rate maximization problem can be stated as
\begin{subequations}
	\begin{align}
		\mathop {\max }\limits_{{{\bf{k}},{\pi _1}, {\pi _2}, {{\bf{\Phi }}_A^{TS}},{{\bf{\Phi }}_B^{TS}},{\bf{u_1}}, {\bf{u_2}}} } \quad& \sum\limits_{t = 1}^T \sum\limits_{m = 1}^M {R_{m,t}^{TS,\sec }} \\
		s.t. \quad & E\left( {{\bf{k}}\left( t \right){{\bf{k}}^H}\left( t \right)} \right) \le {P_{0 }},\forall t, \\
		&{R_{m,t}^{TS}} \ge {R_{0}},\forall t,\\
		&{\bf{SNR}}_{0}^{TS} \ge {\kappa _t},\\
		&(1), (2), (3),
	\end{align}
\end{subequations}
Similarly to problem (16), problem (25) is also a non-convex problem. Hence, finding the optimal solution using the traditional convex optimization is challenging. In the same way, the optimal solutions for $\bf{u_1}$ and $\bf{u_2}$ like (17) can be obtained by 
\begin{equation}
	{{\bf{u_1}}^*} = \frac{{\left( {{{\bf{I}}_{M + L}} \otimes {{\bf{H}}_{s1}}} \right){\bf{\tilde k}}}}{{{{\bf{\tilde k}}^H}\left( {{{\bf{I}}_{M + L}} \otimes {\bf{H}}_{s1}^H{{\bf{H}}_{s1}}} \right){\bf{\tilde k}}}},
\end{equation}

\begin{equation}
	{{\bf{u_2}}^*} = \frac{{\left( {{{\bf{I}}_{M + L}} \otimes {{\bf{H}}_{s2}}} \right){\bf{\tilde k}}}}{{{{\bf{\tilde k}}^H}\left( {{{\bf{I}}_{M + L}} \otimes {\bf{H}}_{s2}^H{{\bf{H}}_{s2}}} \right){\bf{\tilde k}}}},
\end{equation}
where ${\bf{H}}_{s1}={\bf{g}}_{b,s,t}{\bf{g}}_{b,s,t}^H$ and ${\bf{H}}_{s2}={\bf{g}}_{s,t}^{TS}({\bf{g}}_{s,t}^{TS})^H$.

There are typically two primary methods for addressing problem (16) and problem (25) involving continuous action spaces. The first method is to discretize the continuous actions and then utilize a discrete reinforcement learning (RL) algorithm, like Deep Q-Network, to solve them. However, the discretized processing cannot guarantee convergence to the optimal solution of the original problem. Another method is to adopt the policy gradient (PG) algorithm, e.g., reinforce, to directly address the two optimization problem. Nevertheless, the PG algorithms often do not work and desire for the continuous control problems. In order to conquer these challenges, the DDPG algorithm is developed, which can achieve significant results for various continuous control problem\cite{301}.

\section{DRL-BAESD JOINT DESIGN ALGORITHM}
To solve the problem of STAR-RIS-assisted ISAC secure system, we explore DRL-based algorithms, which demand that the complete transfer period follows a Markov Decision Process (MDP). To begin, we denote a sequence of actions and state spaces, transfer probabilities, and instant rewards by \textit{A}, \textit{S}, \textit{P}, and \textit{R}. Then the tuple $<$\textit{A},\textit{S},\textit{P},\textit{R}$>$ is the MDP model for the DRL. For the proposed MDP, the BS and STAR-RIS are treated as agent, and the action taken by the agent, i.e., the BS receive filters and transmit beamforming, along with the STAR-RIS transmitting and reflecting coefficients, are the desired parameters to be optimized\cite{209}\cite{210}. The agent constantly takes actions depending on the immediate environment and then changes the state for the environment in turn so that it can find the optimal reward. By engaging the agent with the environment and a continuous trial-and-error learning process, we can establish the model that motivates the long-term reward to be maximized\cite{207}\cite{208}. 
Meanwhile, the gain of the instant reward $r ^t$ is as shown in Algorithm 1 to satisfy the constraints (16c) and (16d).
\begin{algorithm}
	\renewcommand{\algorithmicrequire}{\textbf{}}
	\caption{Method for evaluating reward}
	\label{alg:1}
	\begin{algorithmic}[1]
		\REQUIRE 
		\quad 
		\IF{${\rm{SNR}}_{0} \le {\kappa _t}$} 
		\STATE $ r ^t={\rm{SNR}}_{0} $
		\ELSE
		\IF{$R_{m,t} \ge R_{0}, \forall m \in M$} 
		\STATE $ r ^t=\kappa _t + M \cdot R_{0} + R_t^{\sec } $
		\ELSE
		\STATE $ r ^t=\kappa _t + \sum\nolimits_{i = 1}^M {\min ({{\rm{R}}_{{\rm{m,t}}}},{{\rm{R}}_0})} $
		
		\ENDIF 
		\ENDIF 		
	\end{algorithmic}  
\end{algorithm}

\subsection{The DDPG Algorithm}\label{AA}
The DDPG algorithm is an offline RL based on the actor-critic (AC) framework, which internally consists of the actor network (AN) and the critic network (CN), both of them iteratively updated following their respective update laws to maximize the accumulated expected reward. By working together under the AC framework, the DDPG algorithm achieves the effective solution of continuous action space problems and enhances the performance in RL tasks.
\begin{figure}[ht]
	\centerline{\includegraphics[width=0.7\textwidth]{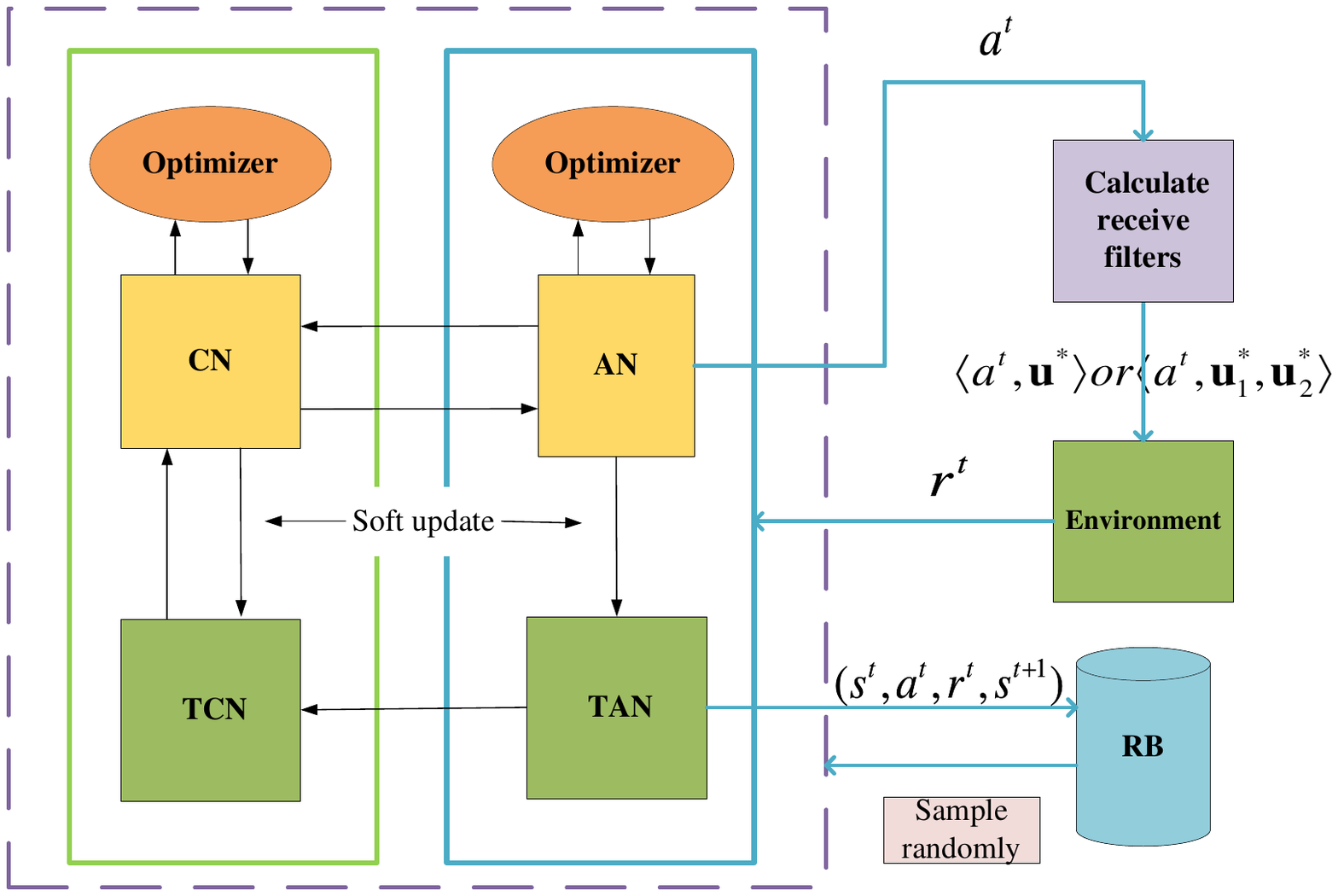}}
	\caption{The network model for DDPG}
	\label{fig}
\end{figure}

\begin{algorithm}[htb]
	\caption{Based on the DDPG co-design algorithm.}
	\label{a.1}
	\textbf{Input:} \quad \leftline{$\bf{CSI}$}
	\textbf{Output:} \quad  \leftline{\{${\bf{k}}$, ${{\bf{\Phi }}_A}$, ${{\bf{\Phi }}_B}$, ${\bf{u}}$\} for problem (16) and \{${\bf{k}},{\pi _1}, {\pi _2}, {{\bf{\Phi }}_A^{TS}},{{\bf{\Phi }}_B^{TS}},{\bf{u_1}}, {\bf{u_2}}$\}} for problem (25) 
	
	\textbf{Initialization:} \quad \leftline{RB, ${\lambda ^\omega }$, ${\lambda ^{\bar \omega }}$, ${\lambda ^\mu }$, ${\lambda ^{\bar \mu }}$} 
	\textbf{Repeat}
	
	\quad \quad Obtaining the starting state $s_0$
	
	\quad  \quad \textbf{Repeat}
	\begin{description}
		\item 1: Get $a^t$ based AN. \;
		\item 2: Obtain new state $s^{t + 1}.$ \;		
		\item 3: ${\bf{u^*}}$ is calculate by (17) or ${\bf{u_1}}^*$, ${\bf{u_2}}^*$ is calculate by (26) and (27). \;		
		\item 4: Obtian the reward $r^{t}$ according to Algorithm 1. \;
		\item 5: Save $\left\{ {a^t,s^t,{r^t},s^{t + 1}} \right\}$ in RB.
        \item 6: {\bf{if}} it's time to update {\bf{then}}
        \item \quad \quad {\bf{for}} each gradient step {\bf{do}}
		\item  \quad \quad \quad      Randomly sampling $D$ tuples 
		\item  \quad \quad \quad Update ${\lambda ^\mu }$ by minimizing (30)
		\item  \quad \quad \quad Update ${\lambda ^\omega }$ by applying (31) 
		\item  \quad \quad \quad Update ${\lambda ^{\bar \omega }}$ and ${\lambda ^{\bar \mu }}$ by (32).		
	\end{description}
	
	\quad \textbf{Until} convergence. \\
	\textbf{Until} convergence.
\end{algorithm}

In Fig. 2, there are the three key features of the DDPG algorithm, as follows\cite{302}:
\begin{enumerate}
	\item Experience replay: The agent stores the acquired experience data $\left( {s,a,r,\bar s} \right)$ in the replay buffer (RB). A batch sampling approach is adopted to randomly pick a group of empirical data from the RB for training when updating the neural network parameters. Such mechanism is helpful to steady training and improve the stability of the algorithm.

	\item Noise exploration: DDPG works with a deterministic strategy, i.e., the actions output by its AN are deterministic and lack proactive exploration for the environment. To increase the explorative properties, a noise $n _{ex}$ is added to the actions output during the training period so that the agent have certain exploration capabilities.  
	\item Target network: With the aim of stabilizing the training and refraining from updating the parameters too frequently, DDPG introduces the target actor network (TAN) and the target critic network (TCN) for the estimation of the target. The parameters of both networks are updated with soft updates so that their parameters are smoothly blended with the current network parameters in a fixed proportion.

\end{enumerate}

The DDPG algorithm update process focuses on optimizing the parameters of AN and CN. For the training process, we randomly sample a batch of experience data $\left( {s,a,r,\bar s} \right)$ from the RB. Subsequently, the TAN is employed to obtain the action $\bar a_i$ in state $\bar s_i$
\begin{equation}
\bar a_i = \bar \omega ({{\bar s_i} \mathord{\left/
		{\vphantom {{\bar s} {{\lambda ^{\bar \omega }}}}} \right.
		\kern-\nulldelimiterspace} {{\lambda ^{\bar \omega }}}}),
\end{equation}
where $\bar \omega ({ \cdot  \mathord{\left/{\vphantom { \cdot  {{\lambda ^{\bar \omega }}}}} \right.\kern-\nulldelimiterspace} {{\lambda ^{\bar \omega }}}})$ is the TAN. And then the TCN $\bar \mu ({ \cdot  \mathord{\left/{\vphantom { \cdot  {{\lambda ^{\bar \mu }}}}} \right.\kern-\nulldelimiterspace} {{\lambda ^{\bar \mu }}}})$ is applied to evaluate the target value of the state-action pair $\left( {s,a} \right)$ :
\begin{equation}
y_i = r + \gamma \bar \mu (\bar s_i,\bar a_i/{\lambda ^{\bar \mu }}),
\end{equation}
where $\gamma  \in \left[ {0,1} \right]$ is the discount factor.

Next, the variance $J$ is minimized between the evaluated value $\mu (s_i,a_i/{\lambda ^\mu })$ of  $\left( {s_i,a_i} \right)$ computed by the CN and the target value by taking advantage of the gradient descent algorithm in order to update the parameters of the CN :
\begin{equation}
J\left(\mu\right) = \frac{1}{D}\sum\limits_{i = 1}^D{\left( {y_i - \mu (s_i,a_i/{\lambda ^\mu })} \right)^2},
\end{equation}
where $D$ is defined as the replay memory allocation.

At the end, the evaluated value for $(s_i, a_i^{new})$, i.e., the accumulated expected reward $Q_i^{new}$, is obtained by utilizing the CN. The parameters of the AN are updated by maximizing $Q_i^{new}$ using a gradient ascent algorithm :
\begin{equation}
J\left( \omega  \right) = \frac{1}{D}\sum\limits_{i = 1}^D\mu \left( {{{{s_i},a_i^{new}} \mathord{\left/{\vphantom {{{s_i},a_i^{new}} {{\lambda ^\mu }}}} \right.
			\kern-\nulldelimiterspace} {{\lambda ^\mu }}}} \right),
\end{equation}
 where $a_i^{new} = \omega \left( {{{{s_i}} \mathord{\left/
 			{\vphantom {{{s_i}} {{\lambda ^\omega }}}} \right.
 			\kern-\nulldelimiterspace} {{\lambda ^\omega }}}} \right)$. 

Furthermore, the TCN and TAN adopt the soft update method, which involves a soft update rate $\varepsilon$ is introduced. Then a weighted average of the old target network parameters and the new corresponding network parameters is done before assigning the value to the TCN, which is given as:
\begin{equation}
{\lambda ^{\bar w}} = \varepsilon {\lambda ^w} + \left( {1 - \varepsilon } \right){\lambda ^{\bar w}}, {\lambda ^{\bar \mu }} = \varepsilon {\lambda ^\mu } + \left( {1 - \varepsilon } \right){\lambda ^{\bar \mu }}.
\end{equation}
On the basis of the previous discussion, the update flow of the joint design algorithm based on DDPG is given in Algorithm 2. 

DDPG is trained to get a deterministic policy $\pi^{*}=\max _{\pi} \sum_{t}\mathbb{E}_{\left(s^{t}, a^{t}\right) \sim \rho_{\pi}}r\left(s^{t}, a^{t}\right)$, i.e., for each state, just one optimal action is considered. The learning goal is to directly maximize the expected value of the accumulated rewards, which may not be effective in tackling complex tasks. Therefore, we next introduce alternative offline policy algorithms, namely SAC. 

\begin{figure}[ht]
	\centerline{\includegraphics[width=0.7\textwidth]{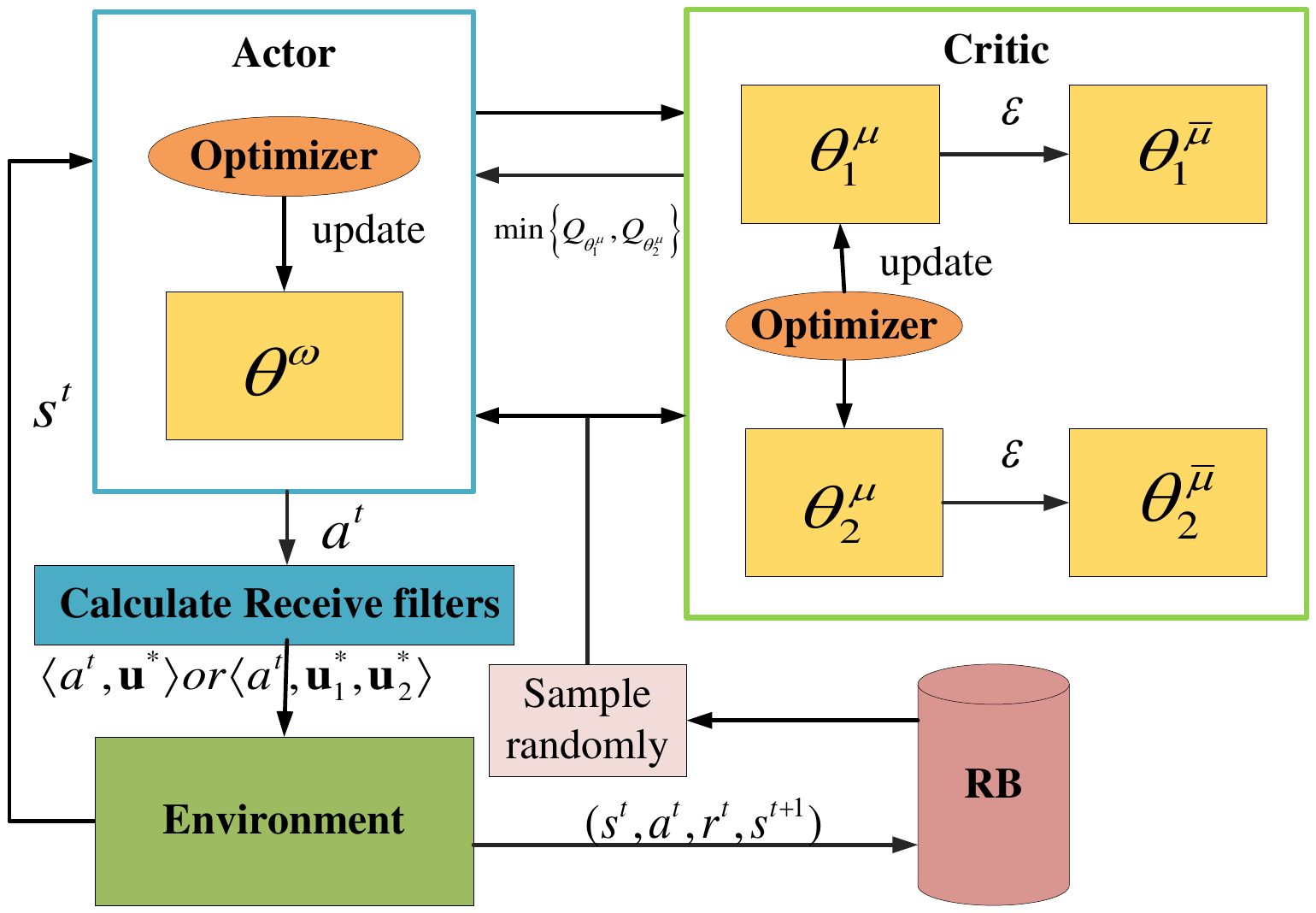}}
	\caption{The network model for SAC}
	\label{fig}
\end{figure}

\subsection{The SAC Algorithm}
\begin{algorithm}[htb]
	\caption{The proposed SAC-based design algorithm.}
	\label{a.1}
	\textbf{Initialization:} \leftline{RB, ${\theta ^\omega }$, $\theta _1^{\mu}$, $\theta _2^{\mu}$, $\theta _1^{\bar \mu}$, $\theta _2^{\bar \mu}$} 
	
	\For{each episode}{
		Obtaining the starting state $s_0$ \\
		\For{each environment step}{
			Obtain ${a^t} \sim {\pi _{\theta ^w}}\left( {{a^t}\left| {{s^t}} \right.} \right)$;\\
			Observe ${s^{t + 1}} \sim p\left( {{s^{t + 1}}\left| {{s^t}} \right.,{a^t}} \right)$;\\
			${\bf{u^*}}$ is calculate by (17) or ${\bf{u_1}}^*$, ${\bf{u_2}}^*$ is calculate by (26) and (27);\\
			Obtian the reward $r^{t}$ according to Algorithm 1;\\
			Save $\left\{ {a^t,s^t,{r^t},s^{t + 1}} \right\}$ in RB.		
		}
		\For{each gradient step}{
			Update $\theta _1^{\mu}$, $\theta _2^{\mu}$ by minimizing (35);\\
			Update ${\theta ^\omega }$ by minimizing (37);\\
			Update $\alpha$ by minimizing (33);\\
			Update $\theta _1^{\bar \mu}$, $\theta _2^{\bar \mu}$ by (32).
		}
		end for
	}
	end for
	
\end{algorithm}

As shown in Fig. 3, the proposed SAC algorithm consists of an AN depicted as $\theta ^w$, two CNs denoted as $\theta _1^{\mu}$, $\theta _2^{\mu}$ and two TCN as $\theta _1^{\bar \mu}$, $\theta _2^{\bar \mu}$. In the following we derive the rules for updating each of these network parameters.

The SAC applies a stochastic policy $\pi^{*}=\max _{\pi} \mathbb{E}_{\left(s^{t}, a^{t}\right) \sim \rho_{\pi}}\left[\sum_{t} r\left(s^{t}, a^{t}\right)+\alpha \mathcal{H}\left(\pi\left(\cdot \mid s^{t}\right)\right)\right]$ contrast to DDPG, which allows the neural network to explore all possible optimal paths, where $\mathcal{H}\left(\pi\left(\cdot \mid s^{t}\right)\right)= - \log \left( {\pi \left( {{ \cdot  \mathord{\left/{\vphantom { \cdot  {{s^t}}}} \right.\kern-\nulldelimiterspace} {{s^t}}}} \right)} \right)$ is the information entropy for $\pi$ and $\alpha$ is the entropy regularization coefficient and is updated by
\begin{equation}
	J\left( a \right) = {\mathbb{E}_{{a^t} \sim {\pi ^t}}}\left[ { - a\log \pi _{{\theta ^\omega }}^t\left( {{{{a^t}} \mathord{\left/
					{\vphantom {{{a^t}} {{\pi ^t}}}} \right.
					\kern-\nulldelimiterspace} {{\pi ^t}}}} \right) - \alpha {H_0}} \right].
\end{equation}
This provides the network with enhanced exploration capabilities and robustness to prevent the policy from prematurely converging to an undesirable local optimum\cite{303}.

First, the CN requires evaluating the Q-value of the current policy and state, which is computed by the equation stated as
\begin{equation}
	Q\left(s^{t}, a^{t}\right) \triangleq r\left(s^{t}, a^{t}\right)+\gamma \mathbb{E}_{s^{t+1} \sim p}\left[V\left(s^{t+1}\right)\right],
\end{equation} 
where $V_{\theta ^{\bar \mu}}\left(s^{t+1}\right)=\mathbb{E}_{a^{t+1} \sim \pi}\left[Q\left(s^{t+1}, a^{t+1}\right)- \alpha \log\right.\left. \pi_{\theta ^w}\left(a^{t+1}\mid  s^{t+1}\right)\right]$ is the expected future reward. And then the soft q-function parameters are updated with minimizing the soft Bellman residuals :
\begin{equation}
J\left( {{\theta ^\mu }} \right) = {{\mathbb{E}}_{\left( {{{s}^t},{{a}^t}} \right) \sim D}}\left[ {\frac{1}{2}{{\left( {{Q_{{\theta ^\mu }}} - \left( {r\left( {{{s}^t},{{a}^t}} \right) + \gamma {{\mathbb{E}}_{{{s}^{t + 1}}\sim p}}\left[ {{V_{{\theta ^{\bar \mu }}}}\left( {{{s}^{t + 1}}} \right)} \right]} \right)} \right)}^2}} \right],
\end{equation} 
where the parameter of the TCN is updated similarly by equation (32).

Under each state, the policy is updated based on the Kullback-Leibler (KL) scatter : 
\begin{equation}
	\! \! 	\pi_{\text {new }}=\arg \min _{\pi^{\prime} \in \Pi} D_{\mathrm{KL}}\left(\pi^{\prime}\left(\cdot \mid s^{t}\right) \| \frac{\exp \left(Q^{\pi_{\text {old }}}\left(s^{t}, \cdot\right)\right)}{Z^{\pi_{\text {old }}}\left(s^{t}\right)}\right),
\end{equation} 
where ${Z^{\pi_{\text {old }}}\left(s^{t}\right)}$ is the distributive function applied to the normalized distribution and can be neglected. And Equation (36) can be optimized with stochastic gradient :
\begin{equation}
J\left( {{\theta ^\omega }} \right) = {\mathbb{E}_{{{s}^t}\sim D}}\left[ {{\mathbb{E}_{{a^t}\sim {\pi _{{\theta ^\omega }}}}}\left[ {a\log \left( {{\pi _{{\theta ^\omega }}}\left( {{{{a^t}} \mathord{\left/
						{\vphantom {{{a^t}} {{{s}^t}}}} \right.
						\kern-\nulldelimiterspace} {{{s}^t}}}} \right)} \right) - {Q_{{\theta ^\mu }}}\left( {{{s}^t},{a^t}} \right)} \right]} \right],
\end{equation} 
Also, the SAC algorithm procedure is stated in Algorithm 3.

\subsection{The Complexity of Algorithm}

Although facing a same formulated problem, the performance and complexity of the two methods for solving the problem varies. SAC network is a type of RL algorithm developed according to the maximum entropy principle. SAC maximises the entropy of the strategy while optimising the strategy with greater cumulative reward. By introducing randomness into the policy, agents are allowed to probe the state space more sufficiently and prevent the policy from trapping prematurely in a localized state space. Also, two Q-value networks are deployed to select the minimum value for preventing Q overestimation, which will undoubtedly increase the complexity of the SAC network. 

The complexity of the DDPG algorithm is decided by the specifications of the hired neural network. Let the AN has $L$ layers, each containing $\mu _l^a$ nodes, the complexity of propagation is expressed as $\sum\limits_{l = 0}^L {\mu _l^a\mu _{l + 1}^a}$. With $\mu _b^a$ representing the number of nodes in the actor network for BN layers, $\mu _r^a$ for ’relu’ layers and $\mu _t^a$ for ’tanh’ layers, base on \cite{901}, The necessary number of floating-point operations is expressed as $3\mu _b^a + 3\mu _r^a + 6\mu _r^a$. By applying the same principles to the CN, the complexity of a single prediction and training step can be formulated as ${\rm O}\left( {\sum\limits_{l = 0}^L {\mu _l^a\mu _{l + 1}^a}  + \sum\limits_{l = 0}^L {\mu _l^c\mu _{l + 1}^c}  + 3\mu _b^a + 3\mu _r^a + 6\mu _r^a + 3\mu _b^c + 3\mu _r^c + 6\mu _r^c} \right)$. Since we usually have ${\mu _l} \ge 5$, the computational complexity can be approximated by ${\rm O}\left( {\sum\limits_{l = 0}^L {\mu _l^a\mu _{l + 1}^a}  + \sum\limits_{l = 0}^L {\mu _l^c\mu _{l + 1}^c}} \right)$. Similarly, we can give the complexity of the SAC algorithm as ${\rm{O}}\left( {\sum\limits_{l = 0}^L {\mu _l^P\mu _{l + 1}^P}  + 2\sum\limits_{l = 0}^L {\mu _l^Q\mu _{l + 1}^Q } } \right)$.

\section{Numerical Results}

In this section, the numerical results are presented to evaluate the system performance of the proposed DDPG and SAC algorithms to solve problem (16) and problem (25), respectively. Additionally, we assume a BS with four antennas, STAR-RIS located at (150m, 150m, 15m) and (0m, 0m, 5m), respectively. The ST is situated in $A$, while the LUs and Eve are positioned in $B$. The location distribution in the STAR-RIS aided ISAC system is shown in Fig. 4. Simulation results are provided to validate the performance of the STAR-RIS-assisted ISAC secure communication system,
the required hyperparameters for the proposed SAC algorithm are detailed in Table I\cite{401}, the proposed DDPG is optimized under the optimal conditions to enable fair comparisons \cite{402}.
\begin{figure}[ht]
	\centerline{\includegraphics[width=0.8\textwidth]{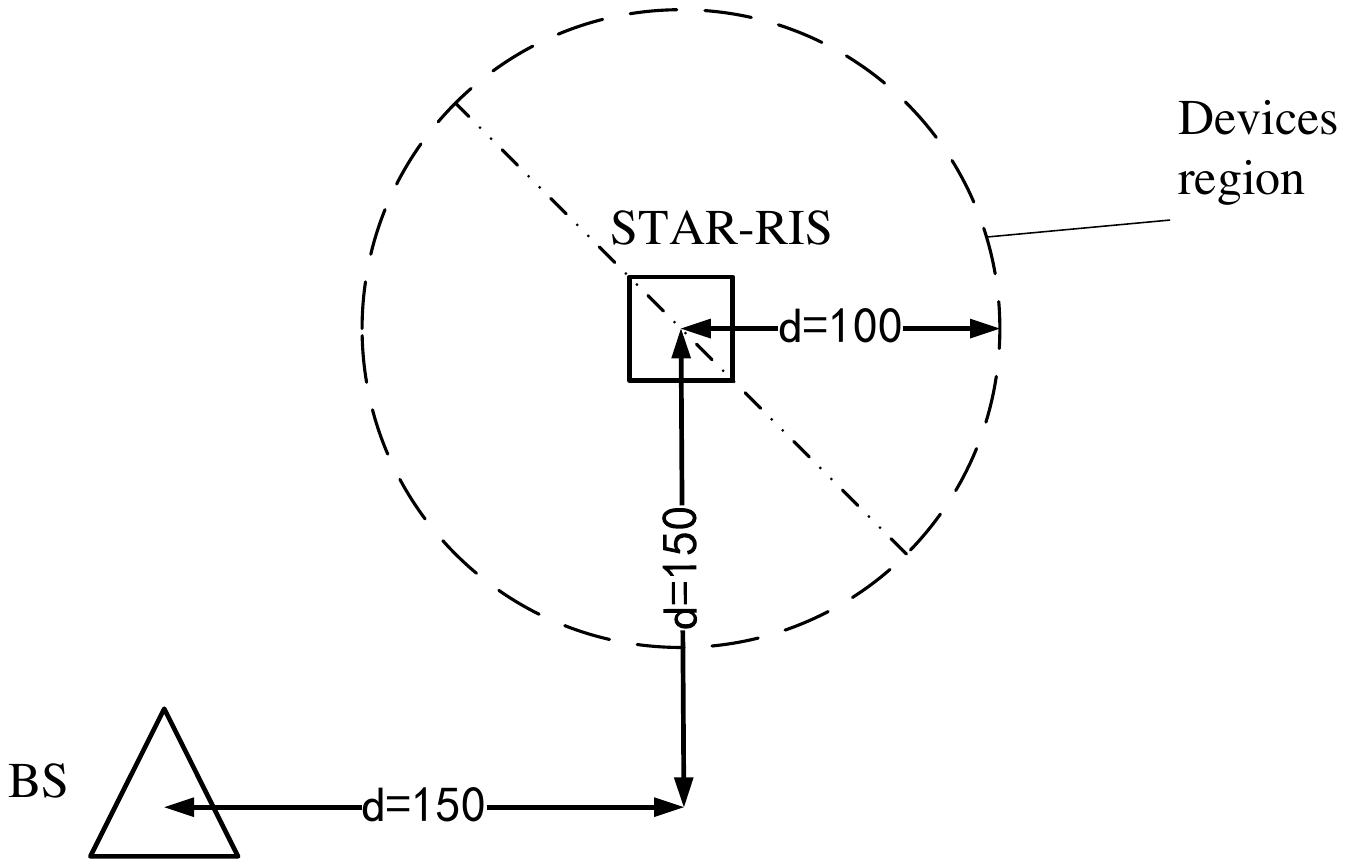}}
	\caption{Location distribution in the STAR-RIS aided ISAC system.}
	\label{fig}
\end{figure}
\begin{table} 
	\centering 
	\caption{Simulation hyperparameter} 
	\label{table_example} 
	
	\begin{tabular}{cccc} 
		
		\toprule 
		\multicolumn{1}{m{2cm}}{\centering Hyperparameter} &  
		\multicolumn{1}{m{3cm}}{\centering Value}&
		\multicolumn{1}{m{2cm}}{\centering Hyperparameter}&
		\multicolumn{1}{m{3cm}}{\centering Value}\\
		
		\midrule 
		Batch siz&64& Target smoothing coefficient&0.0005\\ 
		Layer hidden unit&256&Learning rate&0.0001\\
		Activation function&ReLU& Replay buffer siz&1000000\\
		Number of hidden layers&2& Discount factor&0.99\\
		$R _0$&1 \rm{bps/Hz}&${\kappa _t} $&1~\textrm{dB}\\
		$T$&30& Rician factors&$ 3~\textrm{dB} $ \\
		$P_0$&$ 36~\textrm{dBm} $& Noise power&$ 90~\textrm{dBm} $ \\
        Frequency&$ 2~\textrm{Ghz} $& GPU&$ 3060Ti $ \\
		\bottomrule
	\end{tabular}
\end{table}

In the proposed SAC algorithm, we apply various learning rate to the neural network and study their effects on performance and convergence speed. Fig. 5(a) and Fig. 5(b) shows the rewards obtained by the proposed SAC algorithm for different learning rates with TS and ES model. We can see that a gradual acceleration of convergence as the learning rate increases from 0.00001 to 0.001, which is in line with the general statement that the higher the learning rate, the faster the convergence. For the performance results, the SAC algorithm with a learning rate of 0.001 has the worst performance and the SAC algorithm with a learning rate of 0.0001 has the best performance. This is because too large a learning rate increases oscillations and leads to a sharp drop in performance. In conclusion, the learning rate should be chosen appropriately, neither too large nor too small. In the next simulations, we use 0.001 as the SAC network learning rate used in the simulations regardless of whether STAR-RIS uses the TS or ES protocols. 
\begin{figure*}[t]
	\centering
	\subfigure[]{
		\label{iteration} 
		\includegraphics[width=8cm,height=5.8cm]{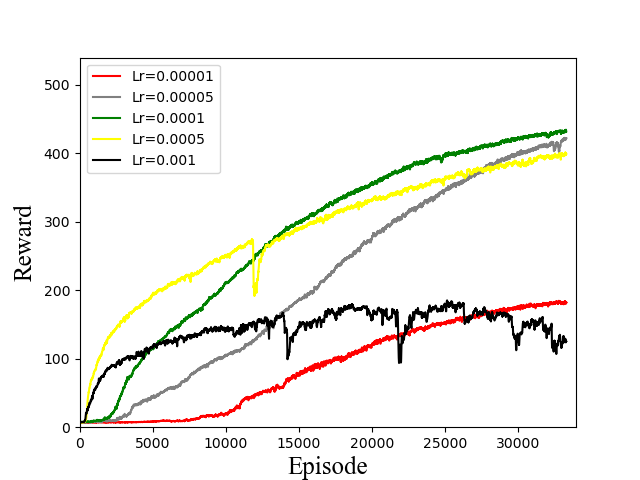}}
	\subfigure[]{
		\label{IP} 
		\includegraphics[width=8cm,height=5.8cm]{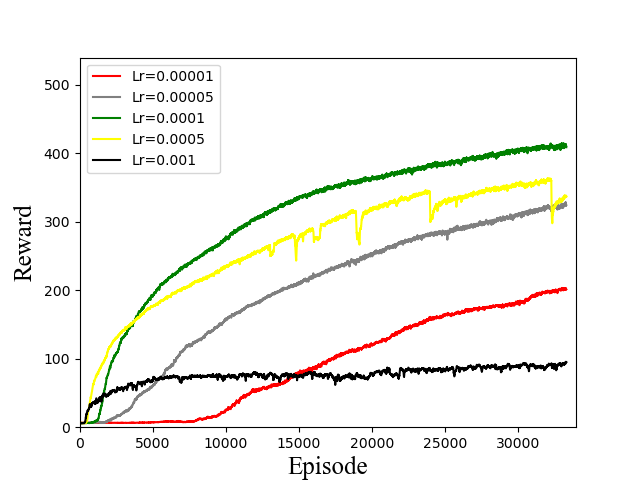}}
	\caption{(a) The return in the ES model. (b) The return in the TS model.}
	\label{fig:subfig} 
\end{figure*}

Figures 6(a) and 6(b) depict the rewards obtained by the DDPG and SAC algorithms for N values of 12 and 24. In this illustration, all schemes demonstrate an increasing trend and eventually reach convergence. Notably, DDPG exhibits a faster convergence rate compared to SAC, and a smaller N leads to quicker convergence. This observation can be attributed to the higher complexity of the SAC network compared to DDPG, and the increase in the number of STAR-RIS elements further contributes to network complexity. Additionally, the SAC algorithm achieves higher rewards than the DDPG algorithm, and a larger number of $N$ effectively enhances reward gains.
 
\begin{figure*}[t]
	\centering
	\subfigure[]{
		\label{iteration} 
		\includegraphics[width=8cm,height=5.8cm]{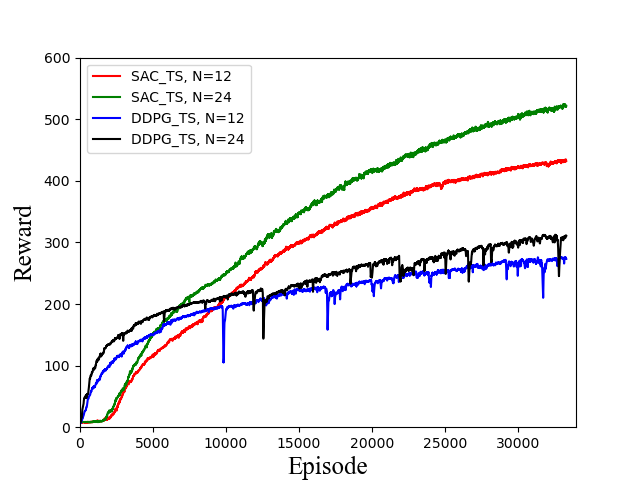}}
	\subfigure[]{
		\label{IP} 
		\includegraphics[width=8cm,height=5.8cm]{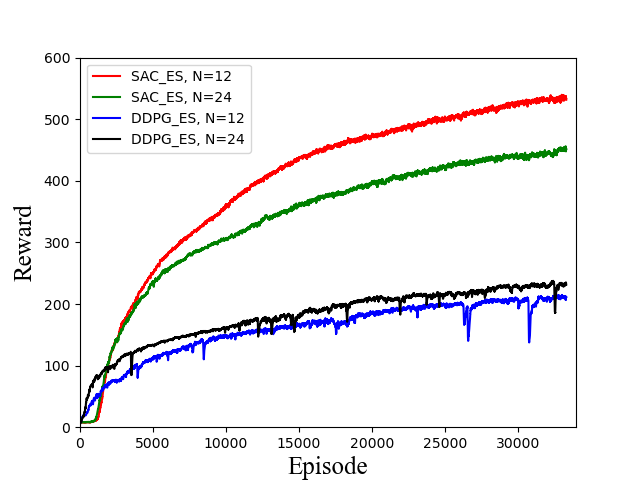}}
	\caption{(a) The return in the ES model with $N$=12 or 24. (b) The return in the TS model with $N$=12 or 24.}
	\label{fig:subfig} 
\end{figure*}

In Fig. 7, we compare the STAR-RIS-assisted ISAC secure system proposed in this work with the following benchmark schemes:
 
\begin{itemize}
	\item The double-spliced RIS: It is constructed by splicing two reflective-only RIS with opposite orientations, and the sum of elements is denoted as $N$.
	\item The conventional RIS: The RIS can only reflect signals rather than transmitting and reflecting at the same time as STAR-RIS, and the sum of elements is denoted as $N$. 
\end{itemize}
To ensure fairness, all three schemes are optimized using the SAC algorithm. According to the results, the average security rate exhibits an increasing trend with $N$, and the STAR-RIS-aidded ISAC outperforms other benchmark scheme. This superiority is attributed to the higher degree of freedom of STAR-RIS elements, allowing for enhanced desired signal strength and interference mitigation. Furthermore, we compare the performance of the two algorithms under ES or TS model. The average secrecy rate obtained by SAC is significantly higher than that of the DDPG algorithm, which is as we expected. The SAC algorithm seeks optimal actions based on maximum entropy has a stronger exploration capability, and the two Q-networks prevent the Q-value from being overestimated and from falling into local optimality.

\begin{figure}[ht]
	\centerline{\includegraphics[width=0.6\textwidth]{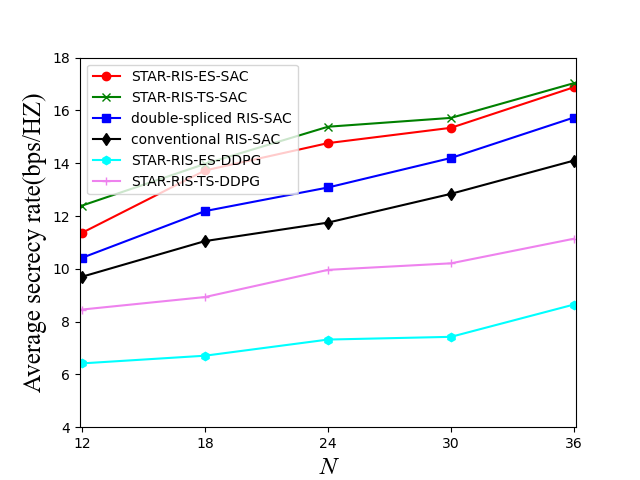}}
	\caption{Average security rate versus $N$.}
	\label{fig}
\end{figure}

In Fig. 8, we illustrate the average security rate and the average running time per episode for both DDPG and SAC as $N$ varing from 10 to 30. The average secrecy rate achieved by both algorithms shows an increasing trend with N, and there is a slight increase in the average time. Additionally, SAC outperforms DDPG in terms of performance, but the average running time is noticeably higher. While SAC exhibits better exploration ability and more effective learning as a reward, it comes at the cost of higher computational complexity compared with DDPG.
\begin{figure}[ht]
	\centerline{\includegraphics[width=0.6\textwidth]{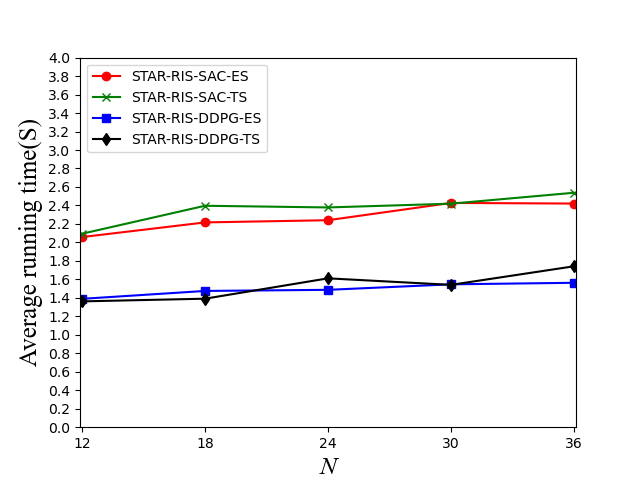}}
	\caption{Comparison of average security rate and running time per
		episode for SAC and DDPG}
	\label{fig}
\end{figure}

In Fig. 9, we present the performance of the two modes of STAR-RIS, i.e., TS and ES, with respect to the variation of the BS power under the two algorithms SAC and DDPG, respectively. From the simulation plot, the average secrecy rate increases with the BS transmit power whether TS or ES. The higher BS power can effectively increase system security. Remarkably, as the power gets higher, this gain becomes less effective.
\begin{figure}[ht]
	\centerline{\includegraphics[width=0.6\textwidth]{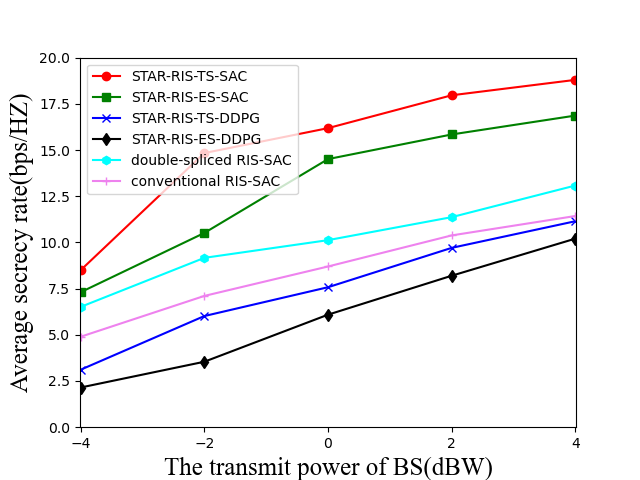}}
	\caption{Comparison of SAC and DDPG-based average security rates and running time}
	\label{fig}
\end{figure}

In addition, we simulate the correlation between the average secrecy sum rate of LUs and the lower bound of the received echo SNR at the BS as shown in Fig. 10. The secrecy rate of LUs decreases with increasing echo SNR in both TS and ES protocal. This is due to the fact that more resources of BS are used to sense the target making the communication less secure. Meanwhile, the TS protocal outperforms the ES protocol at any echo SNR, and the communication security of IUs cannot be guaranteed using the ES protocol when the BS echo SNR is higher than 16.

\begin{figure}[ht]
	\centerline{\includegraphics[width=0.6\textwidth]{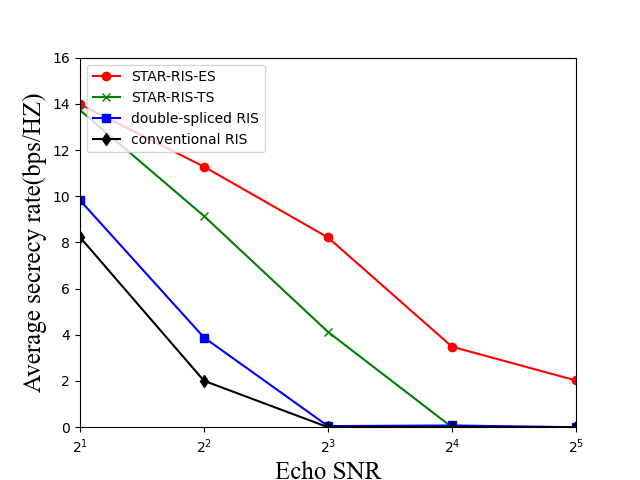}}
	\caption{Average security rate per episode with respect to the upper threshold of echo SNR}
	\label{fig}
\end{figure}

\section{Conclusions}
In this paper, we considered the dual-secure communication issue for the STAR-RIS-assisted ISAC system. The sum secrecy rate of all LU was maximized
with the minimum echo SNR for ST and the achievable rate for LU constraint. The DDPG algorithm and SAC algorithm were proposed to jointly design the receive filters and transmit beamforming of BS, and the transmitting and reflecting coefficients of STAR-RIS. Simulation results demonstrated that the effectiveness of the proposed DRL algorithms and the performance gain of STAR-RIS compared to the conventional RIS.

\bibliographystyle{ieeetr}
\bibliography{refer}

\end{document}